\documentclass[letterpaper,10pt]{article} 
\usepackage{osameet3} %% use version 3 for proper copyright statement

\usepackage{amsmath,amssymb, subfig}
\begin{document}

\title{Phase Compensation for Continuous Variable Quantum Key Distribution}

\author{Hou-Man Chin,$^{1,2*}$ Darko Zibar,$^2$ Nitin Jain,$^1$ Tobias Gehring,$^1$ Ulrik L. Andersen$^1$}
\address{$^1$\mbox{Center for Macroscopic Quantum States, bigQ, Department of Physics, Technical University of Denmark,} 2800 Kongens Lyngby, Denmark\\
$^2$\mbox{Department of Photonics Engineering, Technical University of Denmark,} 2800 Kongens Lyngby, Denmark}
\email{*homch@fysik.dtu.dk}

\begin{abstract}
 The tracking and compensation of phase noise is critical to reducing excess noise for continuous variable quantum key distribution schemes. This work demonstrates the effectiveness of unscented Kalman filter for phase noise compensation. 
\end{abstract}


\section{Introduction}

Current encryption schemes are based on assumptions of intractable mathematical complexity. It may be that due to unforeseen improvements in decryption algorithmic efficiency or a paradigm shift in computation ability such as quantum computing, that these assumptions fail resulting in compromised security. Quantum key distribution (QKD) is a cryptographic scheme that provides mathematically provable security. There are two main flavours of QKD, discrete variable and continuous variable (CV-QKD) \cite{Diamanti2015}. This work concerns Gaussian modulated CV-QKD, or coding the quadratures of the optical field with Gaussian distributed symbols. CV-QKD is typically conducted in a very low signal to noise ratio (SNR) regime. It is also very stringent on the degree of reconstruction required for the received signal after demodulation and digital signal processing.

Due to the very low SNRs associated with the quantum signal, implementing the typical digital signal processing (DSP) methods used in coherent optical telecommunications can be very challenging. One key obstacle is compensation of laser phase noise as the typical quantum signal operates in the region of tens of MBaud leaving it very vulnerable to laser phase noise. Performing DSP on the quantum signal itself is additionally challenging if a Gaussian alphabet is used since constellation points do not simply lie on a ring or set of rings in the IQ plane. 

A standard practice is to insert one or more pilot tones to provide a sufficiently high SNR reference for carrier phase estimation (CPE) and taking its argument \cite{Kleis2017, Laudenbach2019} (requiring considerable optimization) or used an Extended Kalman Filter \cite{Kleis2018} while modulating a discrete modulation format such as M-ary phase shift keying. This work investigates the use of an Unscented Kalman Filter (UKF) on the inserted pilot tone and its impact on the excess noise, assuming a Gaussian alphabet. A lower tolerable pilot SNR allows for wider filtering of the pilot tone accounting for the frequency drift of the lasers, additionally a lower power pilot tone can potentially avoid cross phase modulation from the pilot tone to the quantum signal.

\section{Simulation Setup}
In the simulation, a $2^{15}$ long sequence of symbols is generated from a Gaussian distribution at 50 Mbaud. This sequence is upsampled to 1 Gbaud. A root raised cosine filtered with roll off $\beta = 0.001$ is applied, and the signal is then frequency shifted to 250 MHz. A pilot tone is inserted at 50 MHz frequency offset from the carrier frequency.

The transmitted modulation variance of the quantum signal is set to be 2 shot noise units (SNU) where a SNU is the shot noise power across the equivalent received signal bandwidth. For simplicity thermal noise is not added. The signal is then received by an ideal balanced coherent optical heterodyne receiver. The simulated laser phase noise is generated with a combined laser linewidth of 2 kHz.The laser phase noise is emulated as Wiener process 

The phase noise is estimated using two methods; in the first (standard) method, the pilot signal is band pass filtered and then the phase is estimated by taking the argument of the filtered pilot \cite{Kleis2017}. The other method is to use a UKF to estimate the phase. The phase noise estimates are then used to compensate the phase of the QKD signal after it has been frequency shifted back to baseband and matched filtering applied. To evaluate the performance of the phase noise compensation techniques the excess noise is calculated for the quantum signal. Each simulation is repeated 310 times per pilot SNR and the results averaged.

\section{Results}
\begin{figure*}[t]
\centering
    \subfloat[][]{
        %\hspace{0.8cm}
        \includegraphics[width=0.42\textwidth]{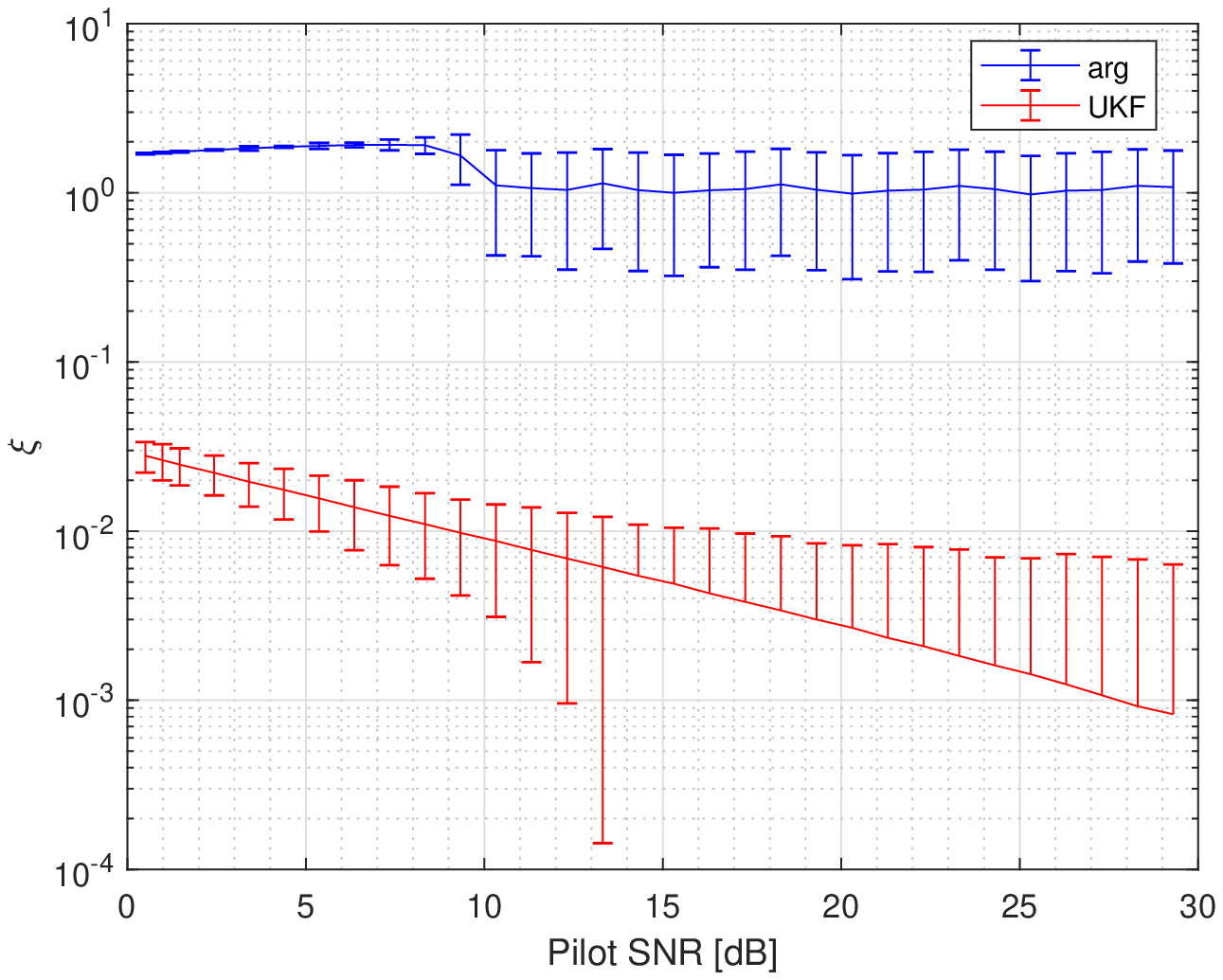}
        \label{fig:EN}}
    \subfloat[][]{
        %\hspace{0.45cm} 
        \includegraphics[width=0.42\textwidth]{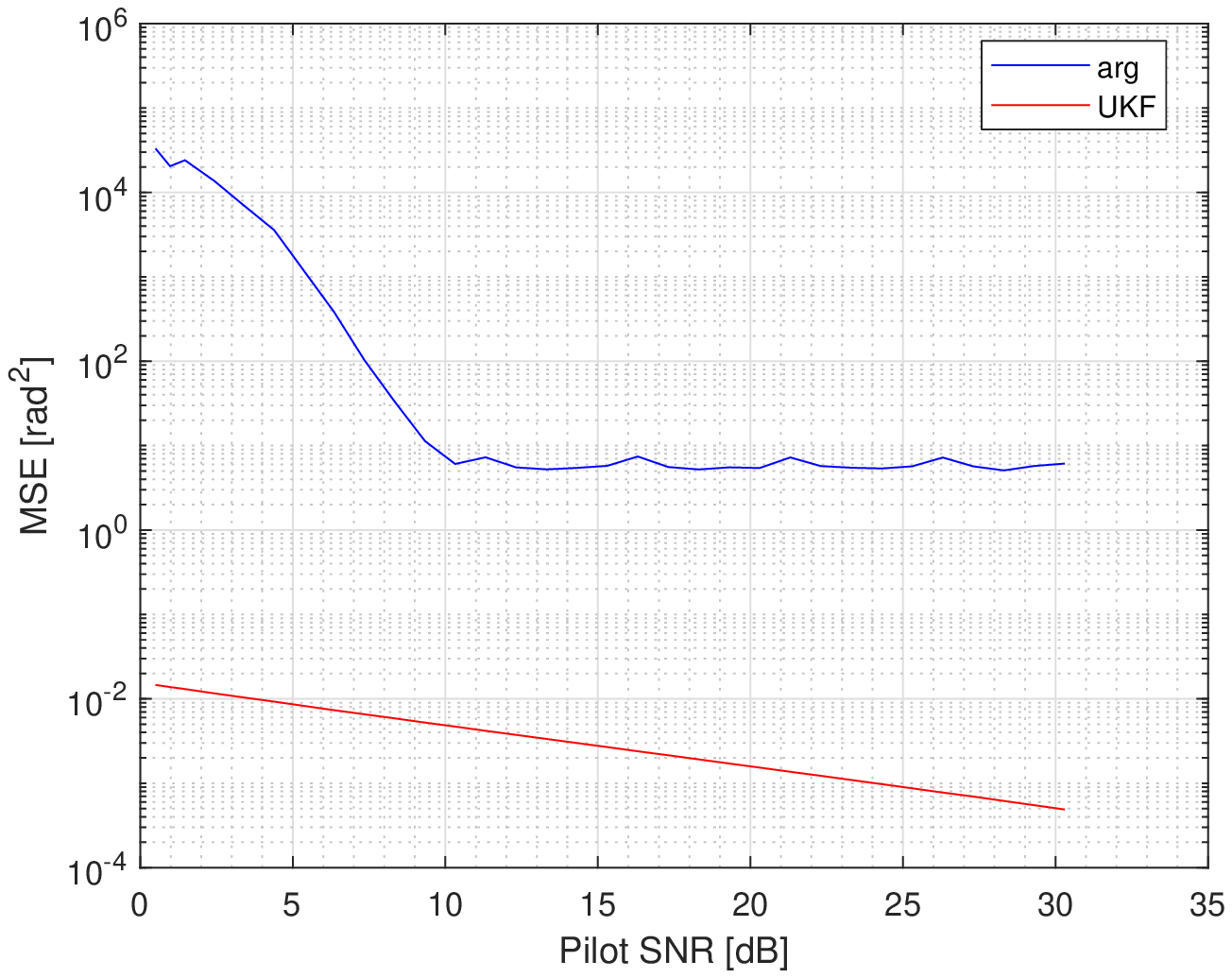}
        \label{fig:MSE}}
    \caption{(a) Excess noise resulting from imperfect phase compensation, (b) Mean squared error of the residual phase after compensation}
    \label{fig:dsp}
\end{figure*}

The metric used to examine the performance of both the argument and the UKF phase noise equalization methods for the QKD channel is excess noise, which is attributed to Eve. Excess noise accumulates on the QKD signal through imperfections which lowers the covariance of Alice's transmitted symbols with Bob's received symbols, and, using an entangling cloner model for the quantum channel, introduces loss in addition to whatever is accounted for during system calibration. The excess noise $\xi$ is calculated using $\xi = y-1-Tx$ where $x = \frac{(I_{A}^{2}+Q_{A}^{2})}{2}$, $y = \frac{(I_{B}^{2}+Q_{B}^{2})}{2}$, $z = \frac{(I_{A}I_{B}+Q_{A}Q_{B})}{2}$, and $T = (\frac{z}{x})^{2}$. I and Q are the in-phase and quadrature components of Alice and Bob's signals.
Figure~\ref{fig:EN} shows the simulated excess noise for the system. It can been seen that the UKF exhibits significantly better performance over the range of pilot SNRs investigated. For typical experimental settings, excess noise from the imperfect phase compensation should be on the order of 0.01 SNUs or lower.

This performance is supported by the mean squared error of the phase estimation methods in Figure~\ref{fig:MSE}. It should be noted that the standard method achieves similarly low excess noise albeit with pilot SNR much greater than shown here.

\section{Conclusions}
This work examined in an extensive set of simulations the performance of an Unscented Kalman Filter based CPE method and compared it to a standard argument based CPE method. It was found that the UKF could enable QKD transmission at pilot SNRs as low as 10 dB while the standard method failed to do so even at 30 dB pilot SNR. Therefore use of machine learning aided DSP is a promising avenue to enable future CV-QKD systems. 

\subsection*{Acknowledgements}
The authors gratefully acknowledge support from Center for Macroscopic Quantum States (bigQ DNRF142), EU project CiViQ (grant agreement no. 820466) and by the European Research Council through the ERC-CoG FRECOM project (grant agreement no. 771878).

\bibliography{20180423.bib}

\end{document}